# Superconductivity in the medium-entropy alloy TiVNbTa with a body-centered cubic structure

*Kuan Li[a#], Xunwu Hu[b#], Ruixin Guo[c,d], Wenrui Jiang[e], Lingyong Zeng[a], Longfu Li[a], Peifeng Yu[a], Kangwang Wang[a], Chao Zhang[a], Shu Guo[c,d], Ruidan Zhong[e], Tao Xie[b], Dao-Xin Yao[b*], Huixia Luo[a*]*

[a] School of Materials Science and Engineering, State Key Laboratory of Optoelectronic Materials and Technologies, Key Lab of Polymer Composite & Functional Materials, Guangdong Provincial Key Laboratory of Magnetoelectric Physics and Devices, Sun Yat-Sen University, No. 135, Xingang Xi Road, Guangzhou, 510275, China

[b] Guangdong Provincial Key Laboratory of Magnetoelectric Physics and Devices, Center for Neutron Science and Technology, School of Physics, Sun Yat-Sen University, Guangzhou, 510275, China

[c] Shenzhen Institute for Quantum Science and Engineering, Southern University of Science and Technology, Shenzhen 518055, China.

[d] International Quantum Academy, Shenzhen 518048, China.

[e] Tsung-Dao Lee Institute & School of Physics and Astronomy, Shanghai Jiao Tong University, Shanghai 200240, China.

[#] K. Li and X. Hu contributed equally to this work.

*Corresponding author/authors complete details (Telephone; E-mail:) (+86)-2039386124; E-mail address: yaodaox@mail.sysu.edu.cn (D. Yao); luohx7@mail.sysu.edu.cn (H. Luo);

**Abstract**

Here we report the TiVNbTa medium-entropy alloy (MEA) superconductor with the mixed $3d$ - $5d$ elements synthesized by an arc-melting method. The TiVNbTa material has a body-centered cubic structure with the cell parameter a = 3.2547(3) Å. The superconducting properties of TiVNbTa were studied by resistivity, magnetic susceptibility, and specific heat measurements. The experimental results show that the bulk superconducting phase transition temperature of TiVNbTa is about 4.65 K, and the upper and lower critical fields are 49.3(4) mT and 5.9(5) T, respectively, which indicates TiVNbTa is a type-II superconductor. First-principles calculations show that the $d$ electrons of Ti, V, Nb, and Ta atoms play a significant role near the Fermi level. The results show TiVNbTa is a traditional s-wave superconductor ($\Delta C_{el}/\gamma T_c$ = 1.60(2), $\lambda_{ep}$ = 0.76(3)).

## I. Introduction

Medium-entropy alloys (MEAs) and high-entropy alloys (HEAs), two unique entropy alloys materials made of three or more elements with equimolar or nearly equimolar ratios, have drawn increasing amounts of attention[1-3]. These multi-component solid solutions possess high mixing entropy ($\Delta S_{mix}$), and the solid solutions with a raised number of constituent elements show an enhanced $\Delta S_{mix}$. $\Delta S_{mix}$ is also employed as a parameter to define HEAs or MEAs: The $\Delta S_{mix}$ of HEAs is greater than or equal to 1.60 R (R denotes the gas constant), and the range of MEAs is between 0.69 and 1.60 R [4-5]. The high $\Delta S_{mix}$ contributes prominently to the free energy ($\Delta G_{mix}$), stabilizing the MEA and HEA phases at ambient temperature. These multi-component alloys typically form hexagonal closed-packing (HCP), body-centered cubic (BCC), or face-centered cubic (FCC) simple structures, where the atoms organize themselves on crystallographic positions of simple lattices with a high degree of the disorder [1-4, 6]. Bcc phases are more stable at a lower valence electron count (VEC) of 6.87, and fcc phases are more stable when the VEC is larger than 8. Between these VEC values, there are both bcc and fcc phases[7-9].

Compared to conventional alloys, these multi-component alloys show preeminent mechanical, thermal, physical, and chemical properties and robust superconductivity against disorder[10-13], high pressure[14-15], weak magnetic impurity[16-17] and extremely strong $s$-wave coupling[18]. Therefore, the preparation and investigation of HEA or MEA superconductors are still of scientific and technological interest. To date, the HEA or MEA superconductors exhibit different dependence on the VEC from those of crystalline and amorphous alloy superconductors. According to Matthias' rule, CsCl-type HEA superconductors have maximum superconducting transition temperatures ($T_c$s) near a VEC of 5.9[19-20]. Recently, a prominent CsCl lattice-type crystal HEA system (ScZrNbTa)$_{1-x}$[RhPd]$_x$, and the optimal composition (ScZrNbTa)$_{0.65}$(RhPd)$_{0.35}$ with VEC = 5.94 yields the maximum $T_c \approx 9.3$ K[20]. Most of the time, hcp HEAs have been linked to a poorer solid solution strength[21]. In contrast, the maximum $T_c$ of superconducting bcc HEAs is reached at VEC = 4.7 [22]. Moreover, using the different combinations of 3*d*, 4*d*, and 5*d* elements is an effective way to adjust the VEC and then search the new HEA or MEA superconductors or improve their $T_c$s.

In this work, we designed the TiVNbTa MEA with VEC = 4.75, which was synthesized using an arc-melting technique. The superconducting properties were studied by combining the experiment and the first-principles calculations. X-ray diffraction (XRD) results suggest that the TiVNbTa compound adopts the bcc structure (space group *Im*-3*m*, No. 229). Analyses of resistance, magnetic susceptibility, and specific heat show that the MEA TiVNbTa is a type-II *s*-wave superconductor with $T_c$ = 4.65 K, $\mu_0H_{c1}(0)$ = 49.3(4) mT, $\mu_0H_{c2}(0)$ = 5.9(5) T, respectively. The electronic structure of TiVNbTa intermetallic compounds is dominated by the *d*-orbital states of transition metal elements, according to the first-principles calculations.

**II. Methods**

The polycrystalline TiVNbTa was synthesized by an arc-melting method using the elements titanium (99 %, 325 mesh, and Alfa Aesar), vanadium (99.5 %, 325 mesh, and Alfa Aesar), niobium (99.8 %, 325 mesh, and Alfa Aesar) and tantalum (99.97 %, 325 mesh, and Alfa Aesar) as raw materials. 300 mg of elemental powders are pressed into lumps in 1:1:1:1, the top layer of tantalum powder, and the bottom layer of other elements. The electric arc furnace was filled and deflated three times, and finally, the arc melting was carried out in 0.5 atmospheres of argon. The alloy was arc-melted in a water-cooled copper furnace. The ingot was remelted at least six times, flipping them between each melt, to create an equal distribution of elements in alloys using this melting procedure. The resulting air-stabilized material was annealed under a vacuum at 1000 °C for five days.

Powder X-ray diffraction (XRD) was applied to confirm the crystal structure. The XRD data were collected from 10 ° to 100 ° with a step width of 0.01° and a constant scan speed of 1 °/min at room temperature using the MiniFlex of Rigaku (Cu Kα1 radiation). The Rietveld approach refined the XRD pattern using the Fullprof suite package software. Energy spectroscopy (EDS) (EVO, Zeiss) and scanning electron microscopy (SEM) were used to analyze the microstructure and element ratios of cast alloys. In addition, resistivity, zero-field-cooling (ZFC) magnetic susceptibility, and specific heat data were collected on quantum design systems (physical property measurement system DynaCool, Quantum Design, Inc). The resistivity was measured using the standard four-probe method.

The experimental lattice structure parameters and atomic positions were adopted in first-principles calculations. The chemically disordered solutions of alloy TiVNbTa were modeled by the "mcsqs" code of the Alloy Theoretic Automated Toolkit (ATAT) [23]. The 2 × 2 × 2 supercell with 16 atoms was built to find the best sqs structure, which most satisfies the correction function of random solutions. According to the Vienna *ab initio* simulation package's (VASP) implementation, we carry out the first-principles calculations[24-25] based on the density functional theory (DFT). For exchange-correlation functions, the Perdew-Burke-Ernzerhof (PBE) generalized gradient approximation (GGA) is utilized[26]. The projector augmented-wave (PAW) potential[27] with a 400 eV plane-wave cutoff energy is employed. The self-consistent calculations employ a $\Gamma$-centered 6 × 6 × 6 k-points mesh within the Monkhorst-Pack framework for Brillouin zone sampling. An energy convergence criterion of $10^{-6}$ eV was given for the electronic self-consistent loop.

Additionally, the systems with strongly correlated (localized) d electrons, like antiferromagnetic transition metal oxide NiO for instance, that need the DFT+U method to improve the description in the DFT calculations. However, the medium-entropy alloy bcc-TiVNbTa is a typical metal with the itinerant d electrons. Therefore, the intensity of the correlation effect of d electrons in the alloy of TiVNbTa should be weak. Thus, we adopt the standard GGA to describe exchange-correlation functions without further correction here. The lattice parameters refined from X-ray diffraction experiment are reliable, in consequence, the experimental lattice structure parameters are fixed in the first-principles calculations.

## III. Results & Discussion

The XRD pattern of the MEA TiVNbTa sample is depicted in **Figure 1a**. From the XRD spectrum, it can be seen that the obtained TiVNbTa is a single phase with no residual impurities. The high level of disorders causes the weak broadening of the peaks.

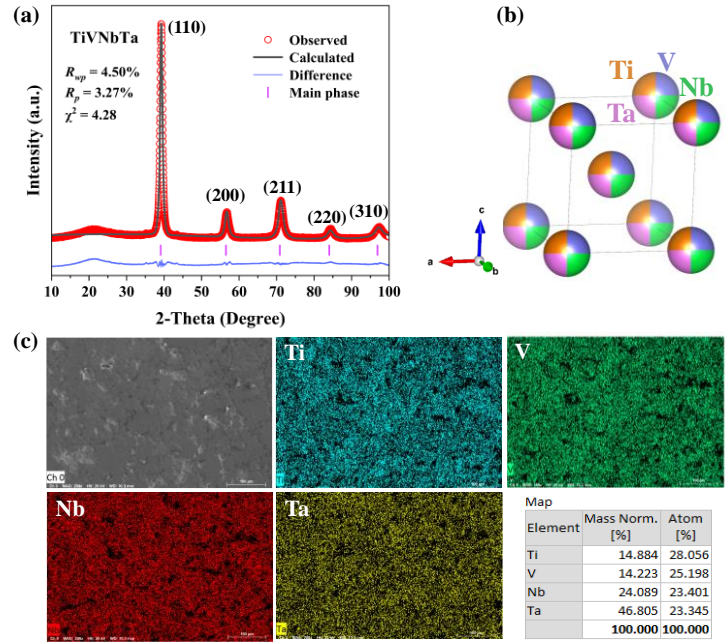

**Figure 1.** Crystal structure characterization of TiVNbTa MEA sample. (a) Rietveld refinement profile of the XRD of the new superconducting phase TiVNbTa MEA. (b) The crystal structure of TiVNbTa with space group *Im*-3*m*. (c) SEM images and EDX elemental mappings for the TiVNbTa MEA.

As shown in **Figure 1b**, the data from the XRD are easily indexed in the bcc structure (*Im*-3*m* space group, and No. 229) with a lattice parameter of $a$ = 3.254(7) Å. When the XRD data were refined using the *Im*-3*m* space group, the $R_{wp}$, $R_p$, and $\chi^2$ parameters were 4.50 %, 3.27 %, and 4.28, respectively. According to the fitting outcomes, the principal phase of the *Im*-3*m* phase accounts for close to 100 %. All observed XRD peaks are labeled with their respective Miller indices. In addition, from SEM-EDX images, we can see that the studied TiVNbTa compounds are microscopically homogeneous. The EDX results give the chemical formula of $Ti_{1.12}V_{1.00}Nb_{0.94}Ta_{0.93}$ **(see Figure 1c)**, which is very close to the design value. These slight deviations can be caused by the limited accuracy of the EDX and the unevenness of the sample surface.

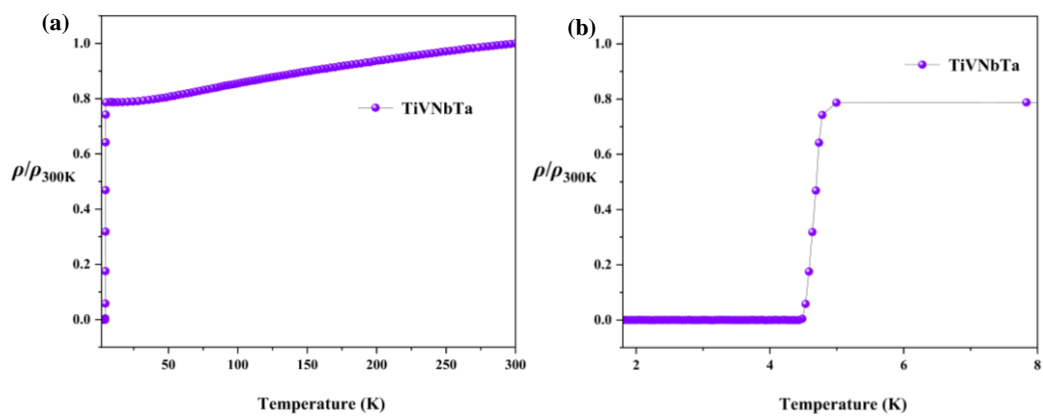

**Figure 2** (a) Temperature dependence of normalized $\rho/\rho_{300K}$ of TiVNbTa MEA sample. (b) Temperature dependence of normalized $\rho/\rho_{300K}$ of TiVNbTa MEA sample at 1.8 - 8 K

The temperature-dependence resistivity of the polycrystalline TiVNbTa MEA sample over a temperature range of 1.8 - 300 K is shown in **Figure 2**. It can be seen that the resistivity of the TiVNbTa sample shows a metallic behavior above $T_c$ (**Figure 2a**). Comparable to those of highly disordered intermetallic or non-stoichiometric ratio compounds, the residual resistivity ratio RRR = $\rho_{(300K)}/\rho_{(8K)} \approx 1.27$ is small. The behavior of low-temperature near $T_c$ is shown in **Figure 2b**. The onset $T_c$ is about 4.99 K, while the zero-resistivity $T_c$ is about 4.48 K. A narrow transition width of 0.51 K indicates our sample is homogenous. Here, we defined the $T_c$ ($T_c \approx 4.65$ K) of the TiVNbTa MEA sample using the midpoint of transition temperatures from the beginning to the end of the superconducting phase transition, which is used to establish the electron phase diagram. **Figure 3a** shows the magnetization data measured under an applied magnetic field of 30 Oe from 1.8 K to 10 K during ZFC. The magnetic susceptibility is calculated by using $\chi_v = \frac{M_v}{H}$ in the formula, $H$ is the applied magnetic field, and $M_v$ is the volume magnetization. The value at which the linear approximation's slope surpasses the normal magnetization's zero value is defined as the critical temperature $T_c$. The strong diamagnetic signal below the critical temperature $T_c$ = 4.5 K proves the presence of superconductivity in the TiVNbTa MEA sample. The shifting trend of $T_c$ is in perfect agreement with the $R - T$ results at a somewhat lower level due to the suppression of the external magnetic field. The $T_c$ values obtained by resistivity and magnetic measurement are consistent. We used the demagnetization factor N computed from the 1.8 K isothermal M(H) data to adjust the magnetization curve further. The shape and orientation of the sample concerning the magnetic field both affect the demagnetization coefficient $N$. For the measured example, the $N$ of the rectangular cuboid sample TiVNbTa MEA sample is about 0.4. The value of $4\pi\chi_v(1-N)$ is close to -1 at 1.8 K, indicating that the volume fraction of Meissner is 100 %. All in all, this suggests that the TiVNbTa MEA sample is a bulk superconductor.

As shown in **Figure 3b**, the low critical field ($\mu_0 H_{c1}(0)$) of the TiVNbTa alloy was further determined by measuring the magnetization M(H) curves at various temperatures. The magnetization measurements of the TiVNbTa MEA sample at different temperatures below $T_c$ vary with the fields. Naturally, magnetization gradually decreases as the magnetic field increases. Low-field linear fitting magnetization data ($M_{fit}$) creates the $M - M_{fit}$ diagram in **Figure 3c**. The magnetization $M$ exhibits a linear relationship with the external magnetic field H when the external magnetic field is weak: $M_{fit} = m + nH$, where $m$ and $n$ are the intercepts and the slope of the line, respectively. Usually, when the difference between $M$ and $M_{fit}$ is around 1 % $M_{max}$, the value of $\mu_0 H_{c1}^*$ is extracted. The points obtained in **Figure 3d** conform to the $\mu_0 H_{c1}^*(T) = \mu_0 H_{c1}^*(0)(1-(T/T_c)^2)$, where $T_c$ is the superconducting transition temperature and $\mu_0 H_{c1}^*(0)$ is the lower critical field at 0 K. As such, the $\mu_0 H_{c1}^*(0)$ of TiVNbTa alloy is estimated to be 49.3(4) mT. Furthermore, when considering the demagnetization factor $N$, $\mu_0H_{c1}(0)$ can be estimated by $\mu_0H_{c1}(0) = \mu_0H_{c1}^*(0)/(1-N)$. The $\mu_0H_{c1}(0)$ of the TiVNbTa MEA sample was calculated to be 821 Oe.

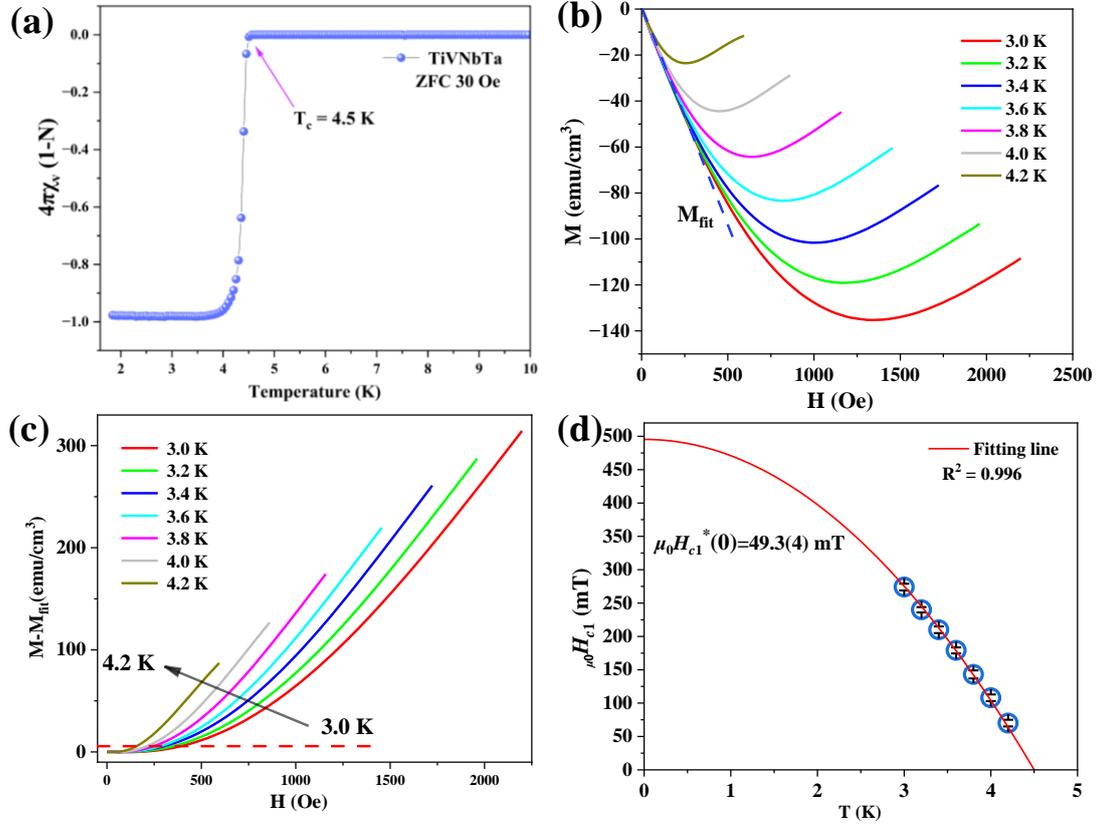

**Figure** 3 (a) The magnetization of ZFC changes with temperature under H = 30 Oe magnetic field and transforms to a diamagnetic Meissner state below 4.5 K. The volume susceptibility $\chi_v$ is corrected by the demagnetization factor measured by M and H. (b) The field-dependent magnetization curve (c) $M - M_{fit}$ curve measured at 3.0 to 4.2 k is used as a function of the magnetic field. (d) Lower critical field estimation.

The study of temperature dependence on resistivity with different fields was carried out to gain insight into the superconducting state. **Figure 4a** shows the resistivity in various magnetic fields. With the increased applied magnetic field, $T_c$ shifts to the lower temperature. In this work, the upper-critical field $\mu_0H_{c2}^*(0)$ is defined by the standard of 50 % normal resistivity and is fitted based on the Ginzberg-Landau (GL) and Werthamer-Helfand-Hohenburg (WHH) theories. **Figure 4b** shows the $H_{c2}$ - T phase diagram of the superconducting state of the TiVNbTa MEA sample. From the linear fitting for TiVNbTa alloy, the resulting slope ($d\mu_0H_{c2}/dT$) is -1.5012(4) T/K. Then, the WHH formula can be used to $\mu_0H_{c2}(0)$: $\mu_0H_{c2}(0) = -0.693T_c(\frac{d\mu_0H_{c2}}{dT})|_{T=T_c}$, which is 4.85 T. It should be emphasized that the derivation from the Pauling limit effect requires that the upper critical field determined by the WHH model must be less than the Pauling limit field $\mu_0H^{Pauli} = 1.86T_c = 8.67$ T. Further, the data are fitted using the GL formula: $\mu_0H_{c2}(T) = \mu_0H_{c2}(0) \times \frac{1-(T/T_c)^2}{(1+(T/T_c)^2}$. Over the whole temperature range, the GL model provided a reasonable fit to the experimental data. The estimated $\mu_0H_{c2}(0)^{GL}$ is 5.9(5) T. For the TiVNbTa MEA sample, WHH's and GL's $\mu_0H_{c2}(0)$s are below and within the Pauli paramagnetic limit.

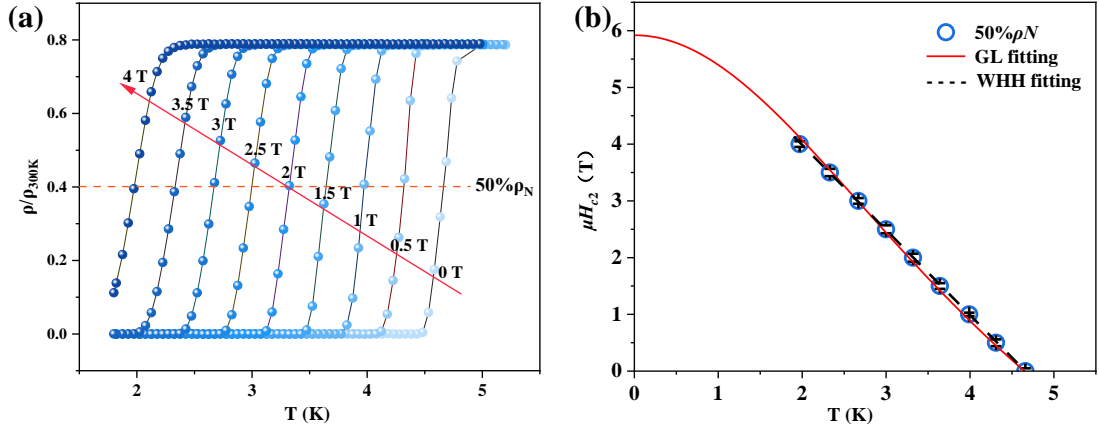

**Figure** 4 (a) The magnetic field dependence of 0 T < $\mu_0H$ < 4 T superconducting transition. The 50 % standard is indicated by a black dotted line (b) The temperature dependence measurement of the upper critical ($\mu_0H_{c2}$) fields for TiVNbTa, the red curve exhibits the refinement by G - L theory. In contrast, the black dotted line displays the refinement by the WHH model.

Several superconducting parameters can be extracted and calculated according to the results of $\mu_0H_{c1}(0) = 49.3(4)$ mT and $\mu_0H_{c2}(0) = 5.9(5)$ T. The GL coherence length ($\xi_{GL}(0)$) of TiVNbTa is estimated to be 74.7 Å by the relationship: $\xi_{GL}^2(0) = \frac{\Phi_0}{2\pi\mu_0H_{c2}(0)}$, where $\Phi_0$ represents the quantum flux (h/2e). The $\xi_{GL}(0)$ value of TiVNbTa MEA is smaller than that of kagome lattice superconductors such as LaIr$_3$Ga$_2$ (84 Å), LaRu$_3$Si$_2$ (107 Å)[28-29], and heavy fermion superconductors (for example, UPd$_2$Al$_3$ (85 Å))[30]. This indicates a stronger electron-electron interaction in TiVNbTa MEA than in the 132-type kagome lattice superconductor and heavy fermion superconductor. In addition, the depth of superconducting penetration ($\lambda_{GL}(0)$) at 0 K can be obtained from $\mu_0 H_{c1}(0) = \frac{\Phi_0}{4\pi\lambda_{GL}^2(0)} \ln\frac{\lambda_{GL}(0)}{\xi_{GL}(0)} + 0.5$ to $\lambda_{GL}(0) = 102$ nm. Thus, from the relation $K_{GL}(0) = \frac{\lambda_{GL}(0)}{\xi_{GL}(0)}$, we get the $K_{GL}(0) = 1.017$. This value is slightly greater than $1/\sqrt{2}$, indicating that TiVNbTa MEA is a type-II superconductor.

Beyond that, to confirm that the observed superconductivity is a property of TiVNbTa, specific heat measurements were carried out in the range of 1.8 to 15 K, as shown in **Figure 5**. The heat capacity is mainly contributed by phonons ($C_{ph.}$) and electrons ($C_{el.}$), which can be described as $\beta T^3$ and $\gamma T$, respectively. The equation $C_p = \gamma T + \beta T^3$ was used to fit the parameter of heat capacity above $T_c$. It is determined that the values of $\gamma$ and $\beta$ are 30.23(3) mJ/mol/K$^{-2}$ and 0.42(2) mJ/mol/K$^{-4}$, separately. The $C_p/T$ vs $T^2$ curves from zero applied fields were calculated in **Figure 5a**. The jump of bulk superconductivity at $T_c = 4.38$ K is sharp, where the obtained $T_c$ is close to those $T_c$s determined from resistivity and susceptibility measurements. In **Figure 5b**, the electron-specific heat ($C_{el.}$) is separated and plotted as $C_{el.}/\gamma T$ against $T/T_c$ after subtracting the phonon contribution. The normalized heat capacity jumps $\Delta C_{el}/\gamma T_c$ is estimated to be 1.60(2) by using the equal area construction method, which is slightly larger than the Bardeen-Cooper-Schrieffer (BCS) weak coupling ratio (1.43), indicating medium coupling in the TiVNbTa MEA alloy. The Debye temperature $\Theta_D$ conforms to the formula $\Theta_D = (12\pi^4 nR/5\beta)^{1/3}$, where n and R represent the number of unit atoms and the molar gas constant in the formula unit, respectively. We can obtain $\Theta_D = 166.54$ K for the TiVNbTa MEA. The electron-phonon coupling constant $\lambda_{ep}$ can be determined to be 0.76(3) on the assumption that $\Theta_D$ and $T_c$ are provided with $\mu^* = 0.13$ by employing the McMillan formula: $\lambda_{ep} = \frac{1.04+\mu^* \ln\left(\frac{\Theta_D}{1.45 T_c}\right)}{(1-1.62\mu^*)\ln\left(\frac{\Theta_D}{1.45 T_c}\right)-1.04}$[31]. In addition, we fit the

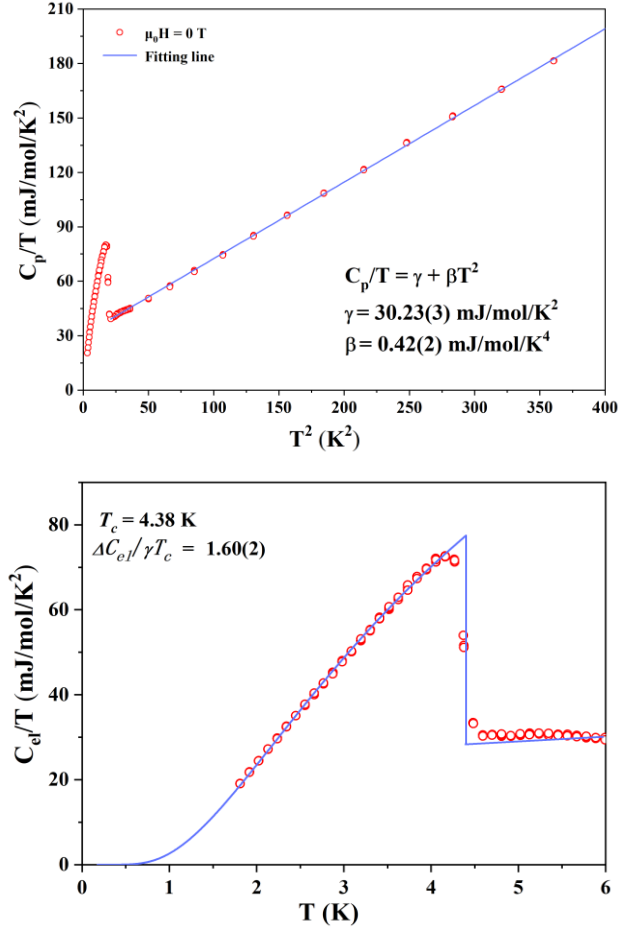

**Figure** 5 The temperature-dependence of specific heat for TiVNbTa MEA sample. (a) The $C_p/T$ vs $T^2$ curves over a temperature range of 1.8 - 20 K, fitted with low-temperature Debye model $C_p/T = \gamma + \beta T^2$. (b) Electronic contribution to the heat capacity divided by temperature ($C_{el}/T$) vs temperature $T$ without an applied magnetic field.

Pauli limiting field ($\mu_0 H^P(T)$) of TiVNbTa by the formula $\mu_0 H^P(T) = 1.86 T_c$. **Table 1** summarizes the relevant superconducting parameters for TiVNbTa, TiHfNbTa[18,] and $Ta_{1/6}Nb_{2/6}Hf_{1/6}Zr_{1/6}Ti_{1/6}$[17]. In addition, it can be seen that using the so-called $\alpha$ model, the traditional $s$-wave gap function can fit the data well and shows that the material is completely gapped[32].

The VEC is a crucial factor impacting the phase stability of crystalline solid solutions in the absence of a significant atomic size influence[33]. It is helpful to report it as the number of valence electrons per atom for MEA and HEA. According to the experimental findings, the MEA and HEA superconductor's VEC is a key factor in influencing the temperature at which it transitions into superconductivity and the stability of the material. The trend of transition metals, which associates the maximum critical temperature with VEC in the transition metal phase, is called the Matthias rule. First-principles calculations show that the TiVNbTa MEA superconductor may be mainly derived from d electrons. Therefore, we summarize the relationship between $T_c$ and VEC of the TiVNbTa MEA superconductor and other MEA and HEA superconductors. The $T_c$ of MEA and HEA superconductor strongly depends on VEC, and the determined critical temperature $T_c$ of MEA and HEA is plotted as a function of valence electron (VE) and atomic ratio *e/a* (red square) in **Figure 6** (red solid line connects the data points). The orange dotted line represents the link between $T_c$ and VEC of crystalline transition metals and their alloys[34-36]. The superconductivity of [TaNb]$_{1-x}$(ZrHfTi)$_x$ is between 4.3 and 4.8. $T_c$ rises as VEC rises and peaks at 7.3 K, close to 4.7 (a pale orange region in **Figure 6**). It is found that the dependence of the critical temperature of (HEA)Al$_x$ on electron counting is between two baselines. However, it is more consistent with the Mathias rule. The addition of aluminum causes a critical temperature dependence closer to crystallization when compared to the trend line of [TaNb]$_{1-x}$(ZrHfTi)$_x$ solid solution. At the same time, the Matthias rule's typical behavior is no longer present in the amorphous transition metal films, where the critical temperature $T_c$ rises monotonically as *e/a* grows[37-38]. It is worth noting that the valence electron dependence of crystalline alloys is apparent. The Matthias rule of crystalline transition metal superconductors' maximum value is found close to *e/a* (*d* electron) = 4.7, a fundamental characteristic of the phenomenon. The VEC value of the MEA alloy superconductor TiVNbTa is 4.75, which follows the Matthias rule, and the superconducting transition temperature reaches 4.65 K. This provides a platform for understanding the relationship between VEC and superconducting transition temperature and gives a practical basis for designing and synthesizing new medium-high entropy superconductor materials.

Table 1. The physical performance of TiVNbTa, TiHfNbTa and $Ta_{1/6}Nb_{2/6}Hf_{1/6}Zr_{1/6}Ti_{1/6}$ superconductor.

|  | TiVNbTa | TiHfNbTa | $Ta_{1/6}Nb_{2/6}Hf_{1/6}Zr_{1/6}Ti_{1/6}$ |
|---|---|---|---|
| $T_c$ (K) | 4.65 | 6.75 | 7.85 |
| $\mu_0 H_{c1}$ (mT) | 49.3(4) | 45.8 | 23 |
| $\mu_0 H_{c2}$ (T) | 4.8(5) | 10.46 | 12.05 |
| $\gamma$ (mJ mol$^{-1}$ K$^{-2}$) | 30.23(3) | 4.70 | 7.45 |
| $\beta$ (mJ mol$^{-1}$ K$^{-4}$) | 0.42(2) | 0.2428 | 0.274 |
| $\Delta C_{el}/\gamma T_c$ | 1.60(2) | 2.88 | 2.22 |
| $\Theta_D$ (K) | 166.54 | 199.9 | 192.3 |
| $\lambda_{ep}$ | 0.76(3) | 0.83 | 0.9 |

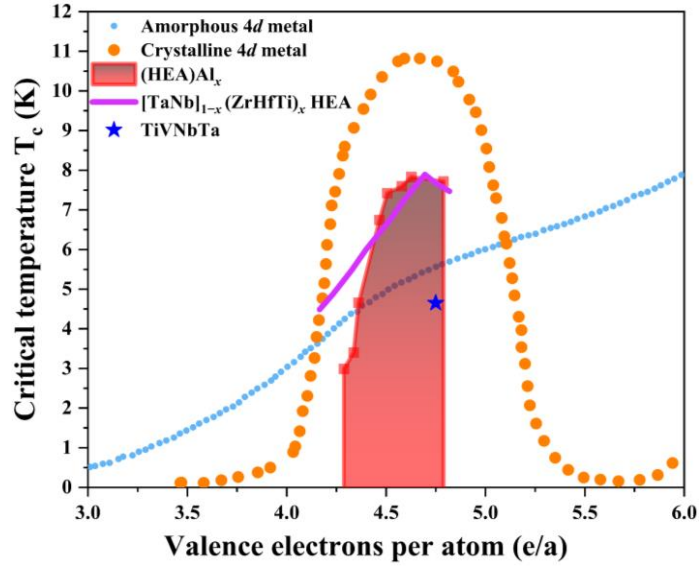

**Figure** 6 Critical temperatures $T_c$ as a function of valence electron count per atom e/a for (HEA)Al$_x$ (red points; the red solid line is a trend line). For comparison the observed trend lines of the transition metals and their alloys in the crystalline form (tangerine dashed line), of the critical temperatures of the HEA solid solution [TaNb]$_{1-x}$(ZrHfTi)$_x$ (purple solid line), and as amorphous vapor deposited films (blue dotted line) are depicted.

As shown in **Figure 7**, the local density of states (DOS) and the partial DOS of *s*, *p*, and *d* orbitals of Ti, V, Nb, and Ta, and the total density of states (TDOS) for alloy TiVNbTa are calculated using DFT. We also consider 24 hypothetically random atomic and structural configurations to examine how the disorder affects the electronic structures characteristics of TiVNbTa alloy. The overall shape of the averaged (24 randoms configurations) and sqs (structure constructed using the mcsqs code) TDOS are almost identical, which means that the disorder of the structure has a tiny effect on the TDOS of alloy TiVNbTa. Therefore, we only consider the sqs structure in the rest of the calculations. As can be seen in **Figure 7a**, the TDOS passing the Fermi level indicates its metallic characteristics. The TDOS is about 2.3 states $eV^{-1}f.u.^{-1}$ at the Fermi level. The local DOS diagram shows that the Ti and V atoms give the largest contribution to TDOS near the Fermi level, Ta and Nb atoms also have some contribution. As shown in **Figure 7(b-e)**, the projected DOS with angular momentum reveals that the d-electrons of all elements have the dominating contribution to the TDOS, i.e., 3*d* for Ti and V, 4*d* for Nb, and 5*d* for Ta, although the p-electron has some contribution. According to these findings, Ti, V, Nb, and Ta's *d*-electrons may be the primary source of superconductivity.

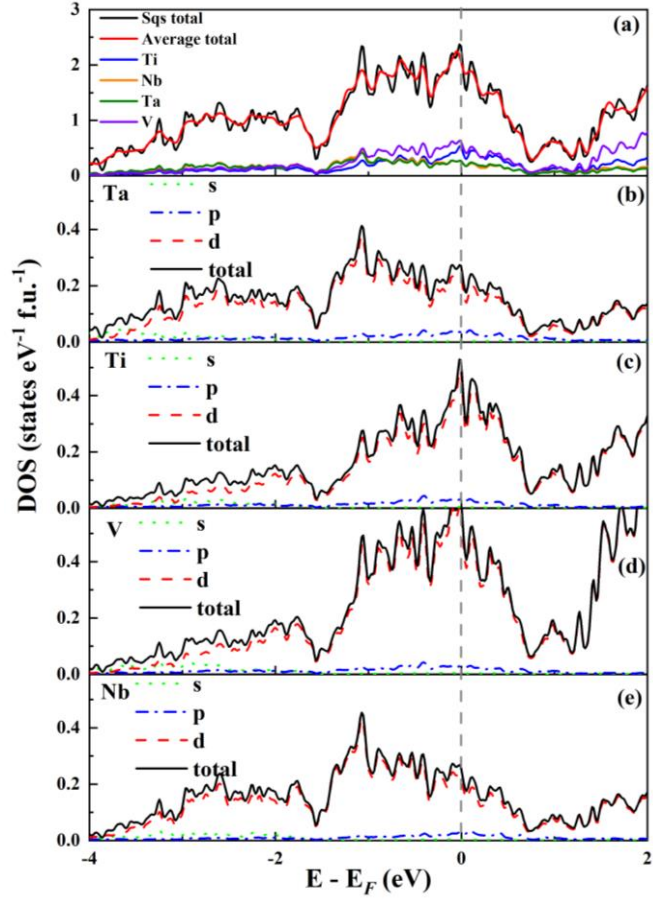

**Figure** 7 (a)Total (black solid line) and local DOS of each element calculated for alloy TiHfNbTa with the structure built by mcsqs code. Average total DOS (red solid line) for 24 assumed random atomic arrangements. (b)-(e) Projected DOS with angular momentum decomposition of each element. The gray dashed line indicates the Fermi level.

**IV. Conclusion**

In summary, we have successfully synthesized the MEA TiVNbTa with bcc crystal structure (*Im*-3*m*) by an arc-melting method. The TiVNbTa alloy is confirmed to be a type II superconductor with $T_c \approx 4.65$ K, $\mu_0 H_{c1} \approx 49.3(4)$ mT, $\mu_0 H_{c2} \approx 5.9(5)$ T, respectively. First-principles calculations show that the *d* electrons of Ti, V, Nb, and Ta play a major role in the formation of superconductivity. Moreover, TiVNbTa MEA is a full-gap superconductor, $\Delta C_{el}/\gamma T_c = 1.60(2)$ and $\lambda_{ep} = 0.76(3)$. In addition, this points the way to synthesize MEA and HEA superconducting materials by Matthias rule.

TiVNbTa provides a new platform to study the unique MEA and superconductivity, which can help to understand the physical mechanism of MEA and HEA superconductors.

**Notes**

The authors declare no competing financial interest

**Acknowledgments**

This work is supported by the National Natural Science Foundation of China (12274471, 11922415), Guangdong Basic and Applied Basic Research Foundation (2022A1515011168, 2019A1515011718), the Key Research & Development Program of Guangdong Province, China (2019B110209003). D. X. Yao and X. Hu are supported by NKRDPC-2022YFA1402802, NKRDPC-2018YFA0306001, NSFC-11974432, NSFC-92165204, Leading Talent Program of Guangdong Special Projects (201626003), and Shenzhen International Quantum Academy (Grant No. SIQA202102). R. Zhong and W. Jiang acknowledges the support for the project by the National key R&D Program of China (Grant No. 2021YFA1401600). T. Xie is supported by Fundamental Research Funds for the Central Universities, Sun Yat-sen University (Grant No. 23qnpy57).